\documentclass{svproc}
\usepackage{url}
\usepackage{graphicx}
\usepackage{sidecap}

\begin{document}
\mainmatter              
\title{Strangeness and light fragment production at
high baryon density}
\titlerunning{Strangeness and light fragment production}  
%
\author{D. Blaschke\inst{1,2,3} \and G. R\"opke\inst{3,4}
\and Yu. Ivanov\inst{2} \and M. Kozhevnikova\inst{2}
\and S. Liebing\inst{5}}
\authorrunning{David Blaschke et al.} 
%
\tocauthor{David Blaschke, Gerd R\"opke, Yuri Ivanov, Marina Kozhevnikova}
\institute{University of Wroclaw, 50-204 Wroclaw, Poland,\\
\email{david.blaschke@uwr.edu.pl},\\ WWW home page:
\texttt{http://www.ift.uni.wroc.pl/$\sim$blaschke}
\and
JINR Dubna, BLTP, Joliot-Curie Str. 6, 141980 Dubna, Russia
\and
NRNU (MEPhI), Kashirskoe Shosse 31, 115409 Moscow, Russia
\and
University of Rostock, Albert-Einstein-Str. 23-24, 18051 Rostock, Germany
\and
TU Bergakademie Freiberg, ITP, Leipziger Str. 23, 09599 Freiberg, Germany
}

\maketitle              

\begin{abstract}
We discuss medium effects on light cluster production in the QCD phase diagram within a generalized Beth-Uhlenbeck (GBU) approach by relating Mott transition lines to those for chemical freeze-out. 
We find that in heavy-ion collisions at highest energies provided by the LHC light cluster abundances should follow the statistical model because of low baryon densities. At low energies in the nuclear fragmentation region, where the freeze-out interferes with the liquid-gas phase transition, selfenergy and Pauli blocking effects are important. 
At intermediate energies the HADES, FAIR and NICA experiments can give new information.
The GBU approach provides new insights to strange hadron production in this energy domain for explaining the "horn" effects. 
\keywords{light clusters, Mott transition, Beth-Uhlenbeck, $K^+$/$\pi^+$ horn effect, ALICE, HADES, NA49, NA61/SHINE, MPD, BM@N}
\end{abstract}

\section{Chemical freeze-out in the QCD phase diagram}

The beam energy scan (BES) programs of heavy-ion collisions (HIC) provide insights into the systematics of particle production under varying conditions of temperature and density of the evolving hadronic fireball created in these experiments.
A remarkable fact is that the thermal statistical model of hadron production makes excellent predictions for particle yields with just two free parameters: the temperature $T_{fo}$ and the baryon chemical potential $\mu_{B, fo}$ at the chemical freeze-out
\cite{Andronic:2017pug}.
The chemical freeze-out concept assumes that 
the system is approximately described by the equilibrium as long as collisions are frequent to establish the corresponding distributions. For an expanding fireball, this is no longer the case at a critical density so that the chemical equilibrium freezes out at the corresponding parameter values for temperature $T$ and baryon number density $n_B$.
An empirical relation for $T_{fo}$ as function of 
$\mu_B$ has been given in \cite{Cleymans:2005xv} 
\begin{equation}
\label{Eq:1}
{T}_{fo}[{\rm GeV}]=0.166-0.139 (\mu_B/{\rm GeV})^2-0.053  (\mu_B/{\rm GeV})^4,
\end{equation}
with 
${\mu_B}[{\rm GeV}]={1.308}/[{1+0.273 \sqrt{s_{NN}}/{\rm GeV}}]$
and  $\sqrt{s_{NN}}=\sqrt{2 m_N E_{\rm lab}+2 m_N^2}$, $m_N=0.939$ GeV.
It has been discussed in \cite{Roepke:2017ohb} that (\ref{Eq:1}) apparently works well down to $T_{fo}\sim 10$ MeV corresponding to HIC at  moderate laboratory energies of $E_{\rm lab} = 35$ AMeV 
\cite{Kowalski:2006ju} which have been analyzed in the quantum statistical freeze-out scheme in 
\cite{Natowitz}, but also up to highest energies provided by heavy-ion collisions at the LHC \cite{Andronic:2017pug}.
 
\begin{table}[!ht]
\begin{center}
	\begin{tabular}{|c|c|c|c|c|c|}
		\hline
		$E_{\rm lab}$ [A GeV] &$\sqrt{s_{NN}}$ [GeV]& $T_{fo}$ [MeV] & $\mu_B$ [MeV]&  n$^{\rm no\, cluster}$[fm$^{-3}$] & n$^{\rm cluster}$[fm$^{-3}$]   \\
		\hline
		1.23& 2.0 & 39.34 & 846.0 & 0.0231 & 0.0529\\
		2.0& 2.35 & 56.38 & 796.8 & 0.0371 & 0.0699\\
		3.85 & 3.0 &  79.956 & 719.1 & 0.0636& 0.0938 \\
		8.0& 4.3 & 108.7 & 601.7 & 0.1045 & 0.1238 \\
		10&4.7&114.7&572.9&0.1122&0.1288\\
		15&5.6&125.0&517.2&0.1225&0.1347\\
		20&6.4&131.8&476.1&0.1260&0.1354\\
		30&7.7&139.6&421.6&0.1251&0.1316\\
		43&9.2&145.7&372.5&0.1196&0.1242\\
		70&11.6&151.8&313.9&0.1086&0.1158\\
		158&17.2&158.6&229.0&0.0875&0.0890\\
		204&19.6&160.0&205.9&0.0813&0.0825\\
		387&27&162.6&156.2&0.0680&0.0687\\
		809&39&164.2&112.3&0.0566&0.0571\\
		2194&64.2&165.3&72.5&0.0470&0.0474\\
		21298&200&165.9&23.5&0.0364&0.0366\\
		\hline
	\end{tabular}
\end{center}
	\caption{\label{Tab:1}
		Freeze-out temperatures $T_{fo}$ and chemical potentials $\mu_B$ according to Eq. (1) by Cleymans et al. \cite{Cleymans:2005xv} for heavy-ion collisions with beam energies $E_{\rm lab}$ in the fixed target mode and the corresponding 
		center of mass energy $\sqrt{s}$ in the nucleon-nucleon system relevant for the collider mode. 
		The last two columns display the total baryon density according to the statistical model without and with light clusters.}
\end{table}
	\vspace{-0.5cm}

In this contribution, we would like to focus on the region between these two extremes, where the transition from baryon stopping to nuclear transparency takes place and the highest baryon densities at freeze-out are reached. 
It is the region of c.m.s. energies of the future NICA facility \cite{Kekelidze:2016hhw}, $\sqrt{s_{NN}}= 2 \dots 11$ GeV, which was partly addressed already by AGS and CERN-SPS experiments as well as the RHIC BES I program and the recent HADES experiment at GSI. 
Besides the MPD and BM@N experiments at NICA it will be covered in future by the low-energy RHIC and the RHIC fixed target programs as well as the FAIR CBM experiment.  
We want to elucidate that the yields of light fragments (clusters) like neutrons (n), protons ({p}), deuterons ($^2$H, d), tritons ($^3$H, t), helions  ($^3$He, h), and $\alpha$-particles  ($^4$He) undergo a change from being strongly affected by medium effects at lowest energies to free-streaming quasiparticle behaviour at highest energies and that a similar transition governs the appearance of the "horn" effects for the ratios of strange to nonstrange hadron production in this energy range.

\begin{figure}[!h] 
	\includegraphics[width=\textwidth]{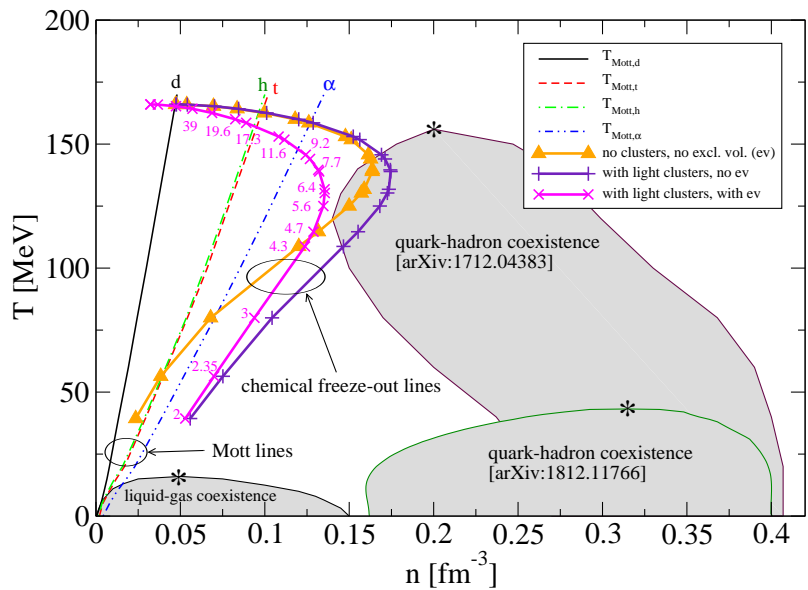}
	\vspace{-1cm}
	\caption{Chemical freezeout lines in the temperature density plane (phase diagram) together with Mott lines for light clusters. The coexistence regions for the nuclear gas-liquid transition and for two examples of the hadron-quark matter transition are shown as grey shaded regions together with their critical endpoints. For details, see text.
	}
 \label{Fig:1} 
 \end{figure}

In Tab.~\ref{Tab:1} we show freeze-out parameter values according to Eq.~(\ref{Eq:1}) for different collision energies from the available and planned beam energy scan programs together with the baryon densities for these parameters, calculated within 
the statistical model with and without light clusters. 
The results are shown by the plus symbols in the QCD phase diagram of Fig.~\ref{Fig:1}, where also the coexistence regions for the nuclear gas-liquid transition and for two examples of the hadron-quark matter transition from Ref.~\cite{Typel:2017vif} and \cite{Bastian:2018mmc} are displayed as grey shaded regions together with their critical endpoints.
\vspace{-1.3cm}

Accounting for an excluded volume of the hadrons according to Ref.~\cite{Albright:2014gva} (version I) results in a reduction of the maximal densities that can be reached at chemical freeze-out (cross symbols), see also \cite{Randrup:2009ch}.
Note that the account of light clusters increases the freeze-out densities at low temperatures by about a factor two
when compared with a result  where no light clusters have been taken into account (triangle symbols) \cite{Randrup:2009ch}. 
The freeze-out at high temperatures proceeds from the quark-gluon plasma (QGP) and basically coincides with the behaviour of the pseudocritical temperatures of the deconfinement transition $T_c$ obtained in lattice QCD simulations 
\cite{Borsanyi:2013bia,Bazavov:2014pvz}.
For $T_{fo}<100$ MeV the freeze-out proceeds from a subphase within hadronic matter, such as the quarkyonic phase characterized by confinement and partial chiral symmetry restoration, resulting in  a triple point in the QCD phase diagram \cite{Andronic:2009gj,Bugaev:2017nir}.

\section{Freeze-out and Mott transition for light clusters}

The particle production measured in heavy ion collisions (HIC) is of interest to infer the properties of dense matter. The time evolution of the fireball produced in HIC demands a nonequilibrium approach to describe the time dependence of the distribution function of the observed products, which are mainly neutrons, protons, and clusters at low energies, but also mesons, hyperons and antiparticles at high energies. Different transport codes have been developed to describe the time evolution of the fireball, but the formation of bound states (clusters) remains an open problem where some semi-empirical assumptions such as the coalescence model are applied.
The freeze-out concept can only be considered as an approximation to describe disassembling matter. It has the advantage that correlations and bound state formation are correctly described within a quantum statistical approach. For a nonequilibrium theory, the equilibrium is a limiting case, and even more, the quasi equilibrium (generalized Gibbs ensemble) serves to define the boundary conditions for the nonequilibrium evolution, see \cite{Zubarev}.

 \begin{figure}[!t] 
	\includegraphics[width=\textwidth]{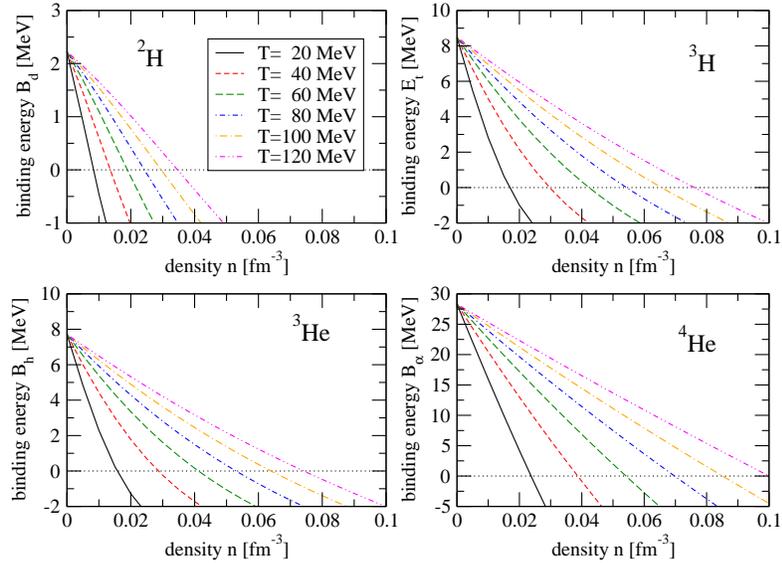}
	\caption{Binding energies of light clusters in nuclear matter as a function of density for different temperatures. Vanishing binding energy defines the Mott transition density. For details of the calculation, see \cite{Typel:2009sy} and 
	\cite{Ropke:2017own}.}
 \label{Fig:2} 
 \end{figure}  

 \begin{figure}[!th] 
	\includegraphics[width=0.64\textwidth]{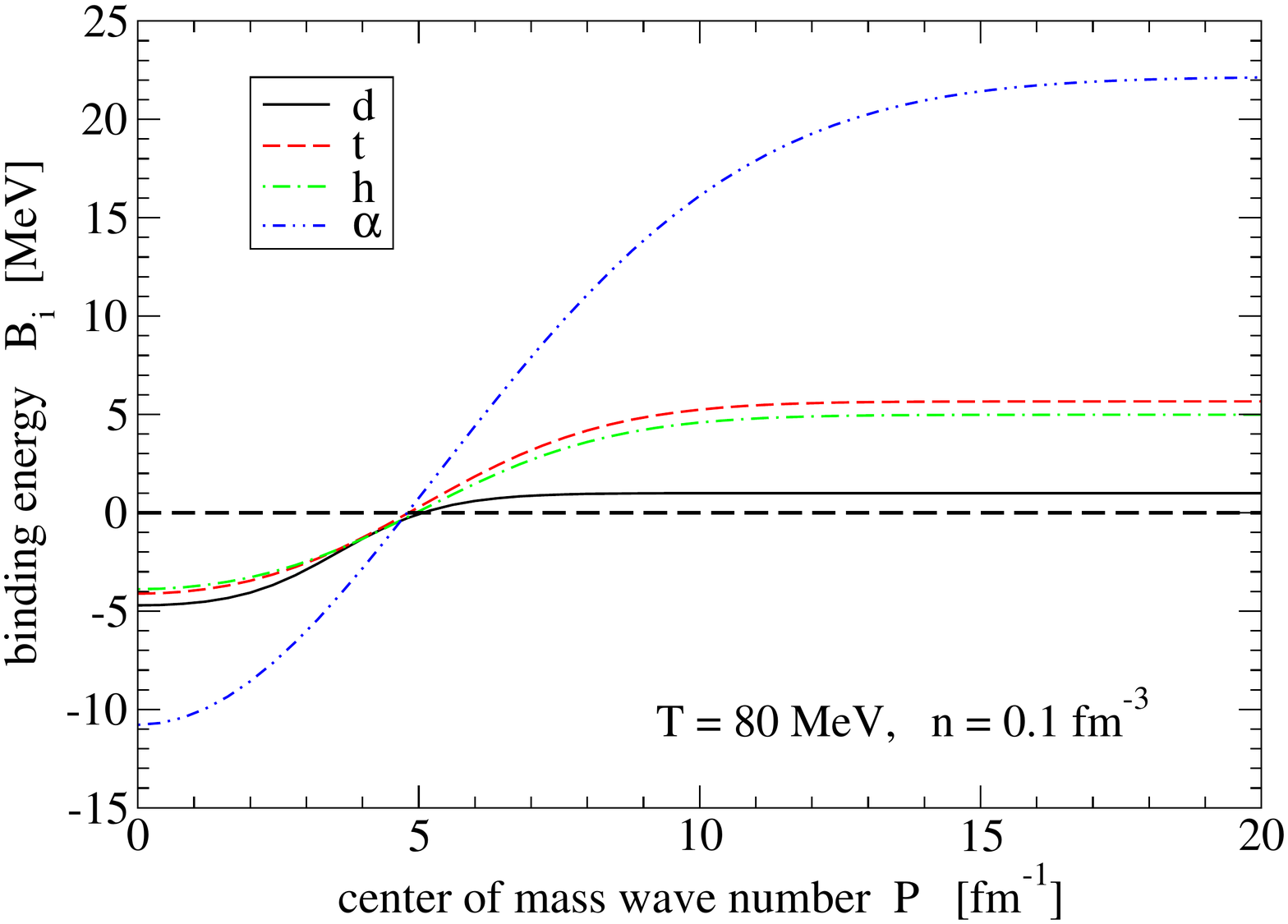}
	\includegraphics[width=0.35\textwidth]{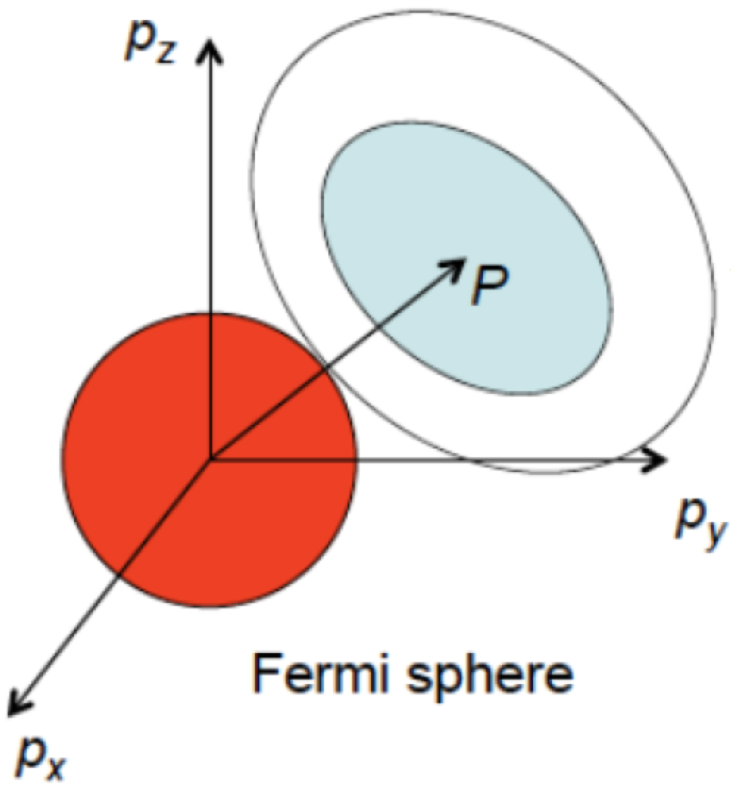}
	\caption{Left: Binding energies of light clusters in nuclear matter as a function of the center of mass momentum $P$ relative to the nuclear medium (Fermi sphere), see right panel. Vanishing binding energy defines the Mott momentum above which the clusters are bound states, see  \cite{Ropke:2017own}.}
 \label{Fig:3} 
 \end{figure}  

A topical question is the debate about the origin of light (anti-)hypernuclear clusters at LHC in the ALICE experiment which follow thermal statistical model predictions with $T_{fo}\sim 156.5$ MeV \cite{Andronic:2017pug} despite the fact that their binding energy is about two orders of magnitude smaller. Are they formed at the hadronization  of the QGP and survive because of the sudden character of the freeze-out or are they formed by coalescence of baryons at a much later stage? 
Here we add the notion of the Mott transition \cite{Ropke:1983lbc} to this discussion. 
Due to an interplay of selfenergy and Pauli-blocking effects in the Bethe-Salpeter equation for light nuclear clusters their binding energy becomes medium dependent, see Fig.~\ref{Fig:2}  and vanishes (Mott effect) along the Mott lines in the phase diagram, see Fig.~\ref{Fig:1}.  For details of the calculation, see \cite{Ropke:2017own}.
To the left of the Mott lines the corresponding cluster is a bound state (at rest in the medium) while to the right of it the cluster is partially dissociated.
For highest energies (and temperatures) the medium does not lead to a Mott  effect and clusters are formed like in free space.
When lowering the energy there is a crossing between freeze-out and Mott lines which strongly depends on details of the calculation as, e.g., the account for a finite momentum of the clusters relative to the medium which would stabilize them, see Fig.~\ref{Fig:3}.  
	The account for light clusters in calculating the freeze-out line removes the second crossing with the Mott lines at low energies. It is very important to study light (hyper)nuclear cluster production in an experiment like HADES, BM@N or MPD
to analyse the transition from high to low $T_{fo}$ systematically. 

\begin{figure}[!th] 
	\includegraphics[width=\textwidth]{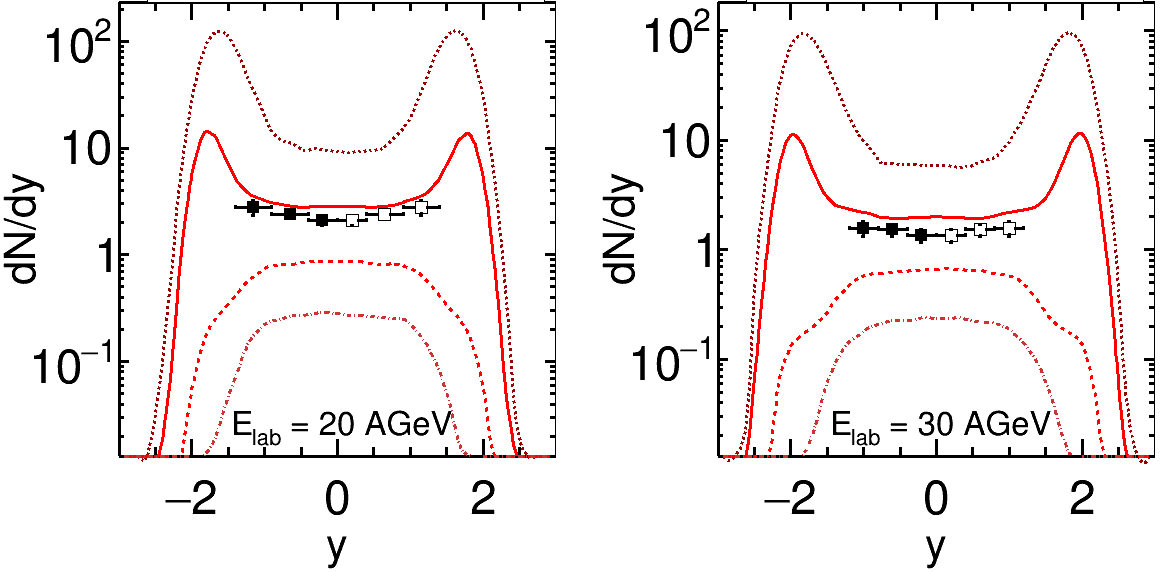}
	\caption{Results of the NA49 collaboration for the deuteron rapidity distribution in Pb+Pb collisions at $E_{\rm lab}=20$ A GeV (left panel) and $E_{\rm lab}=30$ A GeV (right panel) compared with results from the three-fluid hydrodynamics simulation (THESEUS) for impact parameter $b=3$ fm using a crossover equation of state model with a thermal statistical model with (broken lines) and without (solid lines) selfenergy effects.}
 \label{Fig:4} 
 \end{figure}

Here we want to report first results of applying the sudden freeze-out scheme to light cluster formation in the three-fluid hydrodynamics simulation program THESEUS \cite{Batyuk:2016qmb}, by including scalar and vector selfenergies for nucleons as medium effects in the particle distribution functions on the freeze-out surface. 
Calculations for the deuteron rapidity distribution in Pb+Pb collisions at $E_{\rm lab}=20$ A GeV and $30$ A GeV are shown in 
Fig.~\ref{Fig:4} and compared to experimental data from the NA49 collaboration.
We observe that the best description is obtained when the selfenergy effects are discarded. It is necessary to include also the momentum-dependent Pauli-blocking effects in a further development of this study and compare results with recent data from the HADES collaboration at lower energies $E_{\rm lab}=1.23$ A GeV \cite{Szala:2019} 
to find out at which energy the importance of nuclear medium effects for cluster production sets in.
These new results can be compared with previous results for light clusters \cite{Bastian:2016xna} obtained within THESEUS by applying the coalescence scheme \cite{Russkikh:1993ct} for cluster formation in the final state of the collision.
	
 \section{"Horn" effects in strangeness production}

\begin{SCfigure}
	\includegraphics[width=0.65\textwidth]{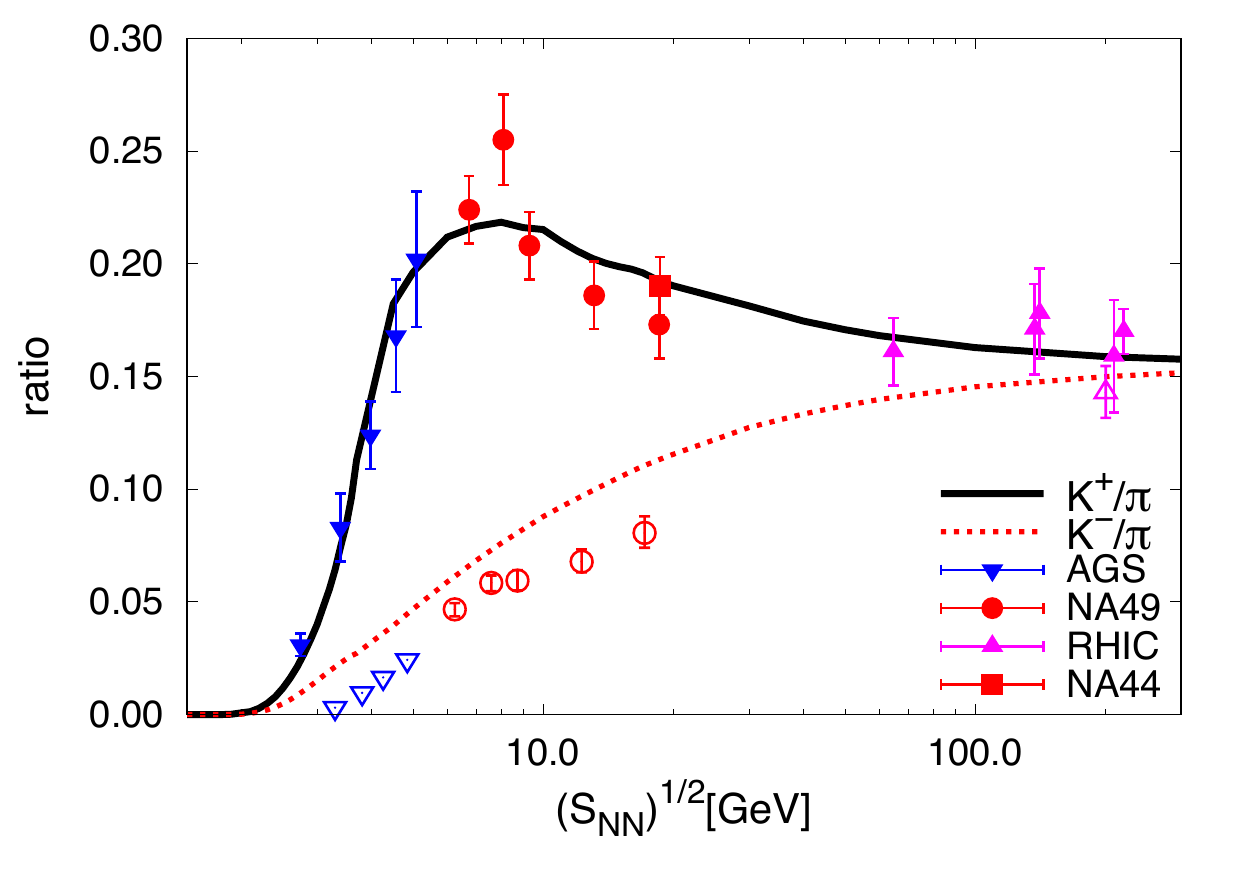}
	\caption{The "horn" in the collision energy dependence of the ratio $K^+/\pi^+$ from different experiments (full symbols) compared to the results from a GBU
	approach to pions and kaons in dense quark matter (solid line)  \cite{Dubinin:2016wvt}. 
	The results for $K^-/\pi^-$ are shown as dotted line, to be compared with the experimental data (open symbols).
	}
 \label{Fig:6} 
 \end{SCfigure}

The effect of a "horn" structure for the ratio of strange to nonstrange particle production as a function of the collision energy has been suggested by Gazdzicki and Gorenstein \cite{Gazdzicki:1998vd} and was established experimentally by the NA49 collaboration \cite{Afanasiev:2002mx} for the particle ratio $K^+/\pi^+$ to be located between the AGS and the SPS domains of collision energy at $\sqrt{s_{NN}}\sim 8$ GeV.
A similar structure has also been found in the energy scan of the ratio $\Lambda/\pi^-$ \cite{Andronic:2017pug}
and they have been attributed to a tricritical point in the QCD phase diagram \cite{Andronic:2009gj,Bugaev:2017nir}. 
Within standard HIC simulations this horn effect has not been reproduced (see, e.g., Ref.~\cite{Batyuk:2016qmb}), but when accounting for effects of partial chiral symmetry restoration on strangeness production already in the hadronic phase  
the horn effect is interpreted as an enhancement of strangeness production on the rising branch of the ratio \cite{Palmese:2016rtq}. 
Another interesting aspect may be the occurrence of a plasmon-like mode in the $K^+$ channel at the hadronisation transition that can be described within a PNJL model for the quark-gluon plasma \cite{Dubinin:2016wvt}, see Fig.~\ref{Fig:6}.
It is also possible that a threshold-like enhancement of the pion production for collisions at $\sqrt{s_{NN}}> 8$ GeV due to the onset of Bose condensation of pions \cite{Begun:2015ifa} contributes to the pronouncedness of the effect \cite{BFKR}. 
A systematic approach should also explain the strong system-size dependence observed by the NA61/SHINE experiment
\cite{Podlaski:2019}.

\vspace{-1cm}

\section{Conclusions} 
We have argued for the importance to combine quantum statistical analyses of low-energy heavy-ion collisions with simulations the NICA parameter region. Light clusters are of interest to determine the parameter values at freeze-out, in particular the density  that is not well known and the role of the in-medium effects, for instance for the deuteron yield.
A second topical goal is the study of strange hadron production in this context, with inputs from chiral quark models of hadrons within the generalized Beth-Uhlenbeck approach in order to pin down the nature of the "horn" effects for strange-to-nonstrange particle ratios.

\subsection*{Acknowledgements}
Discussions with M. Lorenz and M. Szala at the SQM-2019 conference and at the ECT* Trento workshop on "Light clusters in nuclei and nuclear matter"
on the preliminary results of the HADES collaboration are gratefully acknowledged. 
I. Karpenko has provided an update of the particlization routine in the THESEUS programme to include selfenergy effects on 
light clusters.
We thank N.-U. Bastian and S. Typel for providing the data for the coexistence regions in Fig.~\ref{Fig:3} and A. Radzhabov 
for the preparation of Fig.~\ref{Fig:6}.  
S.L. acknowledges the Heisenberg-Landau program for supporting his visit to JINR Dubna.
Y.B.I. was supported by the Russian Foundation for Basic Research, Grants No. 18-02-40084 and No. 18-02-40085 for work related to the upgrade of the three-fluid hydrodynamics code.  
The other work was supported by the Russian Science Foundation under grant No. 17-12-01427.


\begin{thebibliography}{99}

\bibitem{Andronic:2017pug} 
  A.~Andronic, P.~Braun-Munzinger, K.~Redlich and J.~Stachel,
  Nature {\bf 561}, no. 7723, 321 (2018).

\bibitem{Cleymans:2005xv} 
  J.~Cleymans, H.~Oeschler, K.~Redlich and S.~Wheaton,
  Phys.\ Rev.\ C {\bf 73}, 034905 (2006).
  

\bibitem{Roepke:2017ohb} 
  G.~R\"opke, D.~Blaschke, Y.~B.~Ivanov, I.~Karpenko, O.~V.~Rogachevsky and H.~H.~Wolter,
  Phys.\ Part.\ Nucl.\ Lett.\  {\bf 15}, no. 3, 225 (2018).

\bibitem{Kowalski:2006ju} 
S.~Kowalski {\it et al.},
Phys.\ Rev.\ C {\bf 75}, 014601 (2007).


\bibitem{Natowitz}
J. Natowitz {\it et al.},  Phys.\ Rev.\ Lett.\ {\bf 108}, 172701 (2010);
L. Qin {\it et al.},  Phys.\ Rev.\ Lett.\ {\bf 108}, 172701 (2012);
K. Hagel {\it et al.}, Eur. Phys. J. A {\bf 50}, 39 (2014).  
  
\bibitem{Kekelidze:2016hhw} 
  V.~D.~Kekelidze, R.~Lednicky, V.~A.~Matveev, I.~N.~Meshkov, A.~S.~Sorin and G.~V.~Trubnikov,
  Eur.\ Phys.\ J.\ A {\bf 52}, no. 8, 211 (2016).
  
\bibitem{Typel:2017vif} 
  S.~Typel and D.~Blaschke,
  Universe {\bf 4}, no. 2, 32 (2018).
  
\bibitem{Bastian:2018mmc} 
  N.~U.~F.~Bastian and D.~B.~Blaschke,
  arXiv:1812.11766 [nucl-th].


\bibitem{Albright:2014gva} 
  M.~Albright, J.~Kapusta and C.~Young,
  Phys.\ Rev.\ C {\bf 90}, no. 2, 024915 (2014).


  
\bibitem{Randrup:2009ch} 
  J.~Randrup and J.~Cleymans,
  Eur.\ Phys.\ J.\  {\bf 52}, 218 (2016).

\bibitem{Borsanyi:2013bia} 
  S.~Borsanyi, Z.~Fodor, C.~Hoelbling, S.~D.~Katz, S.~Krieg and K.~K.~Szabo,
  Phys.\ Lett.\ B {\bf 730}, 99 (2014).

\bibitem{Bazavov:2014pvz} 
  A.~Bazavov {\it et al.} [HotQCD Collaboration],
  Phys.\ Rev.\ D {\bf 90}, 094503 (2014).

\bibitem{Andronic:2009gj} 
  A.~Andronic {\it et al.},
  Nucl.\ Phys.\ A {\bf 837}, 65 (2010).

\bibitem{Bugaev:2017nir} 
  K.~A.~Bugaev {\it et al.},
  Phys.\ Part.\ Nucl.\ Lett.\  {\bf 15}, no. 3, 210 (2018).

 

\bibitem{Zubarev}
  G.~R\"opke,
  Il Nuovo Cimento C {\bf 39}, 392 (2016).

\bibitem{Ropke:1983lbc} 
  G.~R\"opke, M.~Schmidt, L.~M\"unchow and H.~Schulz,
  Nucl.\ Phys.\ A {\bf 399}, 587 (1983).
  
\bibitem{Typel:2009sy} 
  S.~Typel, G.~R\"opke, T.~Kl\"ahn, D.~Blaschke and H.~H.~Wolter,
  Phys.\ Rev.\ C {\bf 81}, 015803 (2010).
  
\bibitem{Ropke:2017own} 
  G.~R\"opke,
  ``Correlations and Clustering in Dilute Matter,''
in: W.U. Schr\"oder (ed.),
Nuclear Particle Correlations and Cluster Physics
(World Scientific, 2017); 
[arXiv:1703.06734 [nucl-th]],

\bibitem{Gazdzicki:1998vd} 
  M.~Gazdzicki and M.~I.~Gorenstein,
  Acta Phys.\ Polon.\ B {\bf 30}, 2705 (1999).

\bibitem{Afanasiev:2002mx} 
  S.~V.~Afanasiev {\it et al.} [NA49 Collaboration],
  Phys.\ Rev.\ C {\bf 66}, 054902 (2002).

\bibitem{Batyuk:2016qmb} 
  P.~Batyuk {\it et al.},
  Phys.\ Rev.\ C {\bf 94}, 044917 (2016).
  
  \bibitem{Szala:2019}
  M.~Szala (HADES Collaboration), contribution at this conference (2019).
  

\bibitem{Bastian:2016xna} 
N.-U.~Bastian {\it et al.},
Eur.\ Phys.\ J.\ A {\bf 52}, no. 8, 244 (2016).
 
\bibitem{Russkikh:1993ct} 
  V.~N.~Russkikh, Y.~B.~Ivanov, Y.~E.~Pokrovsky and P.~A.~Henning,
  Nucl.\ Phys.\ A {\bf 572}, 749 (1994).

\bibitem{Palmese:2016rtq} 
  A.~Palmese, W.~Cassing, E.~Seifert, T.~Steinert, P.~Moreau and E.~L.~Bratkovskaya,
  Phys.\ Rev.\ C {\bf 94}, no. 4, 044912 (2016).

\bibitem{Dubinin:2016wvt} 
  D.~Blaschke, A.~Dubinin, A.~Radzhabov and A.~Wergieluk,
  Phys.\ Rev.\ D {\bf 96}, no. 9, 094008 (2017).
  
\bibitem{Begun:2015ifa} 
  V.~Begun and W.~Florkowski,
  Phys.\ Rev.\ C {\bf 91}, 054909 (2015).

\bibitem{BFKR}
  D.~Blaschke, A.V.~Friesen, Yu.L.~Kalinovsky and A.~Radzhabov,
  arXiv:1912.13162 [hep-ph].  
  
\bibitem{Podlaski:2019}  
 P.~Podlaski (NA61/SHINE Collaboration),  contribution at this conference (2019).
 
\end{thebibliography}
\end{document}